\documentclass[lettersize,conference]{IEEEtran}
\usepackage{amsmath,amsfonts}
\usepackage{algorithmic}
\usepackage{array}
\usepackage[caption=false,font=normalsize,labelfont=sf,textfont=sf]{subfig}
\usepackage{textcomp}
\usepackage{stfloats}
\usepackage{url}
\usepackage{verbatim}
\usepackage{graphicx}
\usepackage{cite}
\usepackage{amsmath,amssymb,amsfonts}
\usepackage[ruled,linesnumbered]{algorithm2e} % Enables the writing of pseudo code.
\usepackage{tikz,pgfplots}

\usetikzlibrary{positioning}
\usepackage{graphicx}
\usepackage{tabularx}
\usepackage{textcomp}
\def\BibTeX{{\rm B\kern-.05em{\sc i\kern-.025em b}\kern-.08em
    T\kern-.1667em\lower.7ex\hbox{E}\kern-.125emX}}

\definecolor{burlywood}{rgb}{0.87, 0.72, 0.53}

\begin{document}

\title{Blockchain-based Zero Trust on the Edge}

 \author{
 Cem Bicer, Ilir Murturi, Praveen~Kumar~Donta, and Schahram Dustdar\\[0.2cm]

 \IEEEauthorblockA{{Distributed Systems Group}, {TU Wien}, Vienna 1040, Austria.\\   
      }
 }  
%\author{Cem Bicer, Ilir Murturi,  Praveen Kumar Donta, and Schahram Dustdar,
        % <-this % stops a space
%\thanks{I. Murturi is the Corresponding author}
%\thanks{Authors are with Distributed Systems Group, TU Wien, 1040, Austria}
%\thanks{Email: \{cem.bicer, imurturi, pdonta, dustdar\}@dsg.tuwien.ac.at}% <-this % stops a space
%}

\markboth{IEEE Transactions on Consumer Electronics,~Vol.~XX, No.~X, August~2022}%
{Shell \MakeLowercase{\textit{et al.}}: A Sample Article Using IEEEtran.cls for IEEE Journals}


\maketitle

\begin{abstract}
Internet of Things (IoT) devices pose significant security challenges due to their heterogeneity (i.e., hardware and software) and vulnerability to extensive attack surfaces. Today's conventional perimeter-based systems use credential-based authentication (e.g., username/password, certificates, etc.) to decide whether an actor can access a network. However, the verification process occurs only at the system's perimeter because most IoT devices lack robust security measures due to their limited hardware and software capabilities, making them highly vulnerable. Therefore, this paper proposes a novel approach based on Zero Trust Architecture (ZTA) extended with blockchain to further enhance security. The blockchain component serves as an immutable database for storing users' requests and is used to verify trustworthiness by analyzing and identifying potentially malicious user activities. We discuss the framework, processes of the approach, and the experiments carried out on a testbed to validate its feasibility and applicability in the smart city context. Lastly,  the evaluation focuses on non-functional properties such as performance, scalability, and complexity.

\end{abstract}


\section{Introduction}
Traditional cybersecurity methods for computer networks primarily rely on perimeter-based security models. These models emphasize protecting resources through identity verification mechanisms, typically employing cryptography to grant access exclusively to authorized entities. Actors, whether users or devices, establish their identity by presenting login credentials, such as usernames and passwords, or certificates containing relevant details \cite{syed2022zero}. A logical network component validates peers' authenticity before granting access to the requested resource. This authentication approach has historically been effective and remains prevalent in secure networks today.

The rise of the IoT emerged new challenges regarding securing resources \cite{sedlak2023privacy}. Computing continuum infrastructures typically have many heterogeneous devices that all communicate with each other \cite{computers12100198}. The perimeter-based approach faces some limitations in these extensive scenarios, such as ignoring insider threats within authenticated networks and challenges to applying the perimeter concept across the highly dynamic computing continuum infrastructures. As an alternative, Zero Trust Architecture (ZTA) offers a promising solution, emphasizing perimeter-less, continuous verification to protect digital assets from potential threats \cite{rose2020zero}. The core tenet of ZTA is "\textit{never trust, always verify}," advocating rigorous monitoring and verification of all network traffic before granting access to a network or resource \cite{rose2020zero}. Nevertheless, implementing ZTA in the computing continuum infrastructures requires further advanced mechanisms to improve decision-making and security posture.

A smart city is an example of a computing continuum infrastructure with many heterogeneous devices. In a smart city, devices may be equipped with various sensors that monitor and sense the environment. These devices include cameras for monitoring pedestrian crossings and traffic congestion, temperature and humidity sensors for weather monitoring in public parks, sensors for identifying technical issues in mall elevators, etc. These devices communicate directly with the cloud or through an edge server, which transmits the data to the cloud. In the context of a smart city, sub-networks may exist, grouping certain devices or servers to function as independent networks or all public devices could constitute a single extensive network (e.g., city scale). With such many devices operating in a smart city, the potential attack surface becomes significantly large, elevating the risk of system vulnerabilities. Furthermore, if an attacker successfully compromises a device and gains access to the smart city's internal infrastructure, it could significantly impact the smart city's security \cite{gharaibeh2017smart}. 

The ZT approach can help to reduce the above-mentioned security risks. ZT principles focus on protecting \textit{resources} such as data or services, rather than preserving an entire network or domain. With the ZT approach, no implicit trust is assumed, and all entities are treated as untrustworthy at any time by default, whether internal or external. On each incoming request, the ZT system verifies some properties of the requester to decide if the access is granted or rejected. If the access is granted, the given access rights are always as strict and atomic as possible only to allow execution of this specific request. This approach assumes that a connected peer could be compromised at any time at any transaction and checks its privileges, access rights, and previous behavior on every transaction \cite{rose2020zero}. Such a system must ensure the integrity of actors' request history and protection against potential attackers' manipulation of the request validation process. A distributed ledger can mitigate the risk of data tampering while preventing compromised or inactive validation nodes from making decisions on access requests \cite{salman2018security}. This can be achieved through a consensus mechanism involving all validating nodes. 

Therefore, this paper proposes a novel approach based on ZT architecture which integrates blockchain technology into ZT to further enhance the system's security posture. Essentially, the blockchain component serves as an immutable database for saving request history, which is used for verifying the trustworthiness of actors via a consensus. The ZTA components on the edge are responsible for enforcing the defined policies and granting or rejecting incoming requests. We present a Proof-of-Concept (PoC) system and evaluate non-functional properties by executing several test cases. The evaluation focuses on non-functional properties like performance, scalability, and complexity. The results are compared between the various system configurations, and we demonstrate how design decisions may affect the non-functional properties. Our evaluation shows promising results and the feasibility of integrating blockchain in ZT.

The remaining sections are structured as follows. Related work is presented in \textit{Section}~\ref{rw}. \textit{Section}~\ref{section2} introduces and explains our proposed blockchain-based zero trust framework designed for distributed computing continuum systems. Furthermore, we present the implemented PoC by listing its components with
their responsibilities. Evaluation and results are discussed in \textit{Section}~\ref{evaluation1}. Finally, we conclude our discussion in \textit{Section}~\ref{conclusion}.

\section{Related Work}
\label{rw}
Xiaojian et al. \cite{xiaojian2021power} introduced ZT security in a power IoT network. Their system evaluates multiple actors' attributes to calculate trust levels and compares them with resource security levels, which are determined by resource importance. % Actor trust levels are session-based, enabling dynamic access decisions. A central rule base, containing policies, guides trust calculations and adjusts dynamically based on actor behavior, attributes like version and vulnerability patch, and identity. This process is overseen by a central service known as the 'trust analysis engine,' also known as 'Policy Engine' component. 
DeCusatis et al. \cite{decusatis2016implementing} applied the ZT security concept uniquely in a cloud computing context. Their innovative architecture uses steganography to embed authentication data in TCP packets, preventing network fingerprinting. % Authentication data in the transport layer ensures that a TCP connection won't even be established if an actor is untrustworthy. %The system involves two gateways responsible for policy enforcement: the first gateway embeds an authentication token in the initial TCP packet, while the second gateway reads and enforces policies. If an actor is trusted, the second gateway allows the TCP connection; otherwise, it denies the connection, ensuring low latency and high bandwidth without inspecting packet content. 
Samaniego et al. \cite{samaniego2018zero} introduced Amatista, a blockchain-based middleware for IoT resource access management, following the zero trust paradigm. Amatista features a unique two-level hierarchical mining process, with first-level miners validating sender identities and forwarding data to second-level miners. %The latter assesses device access rights and transaction validity using a consensus algorithm before storing transactions in the blockchain. 
Amatista falls short of implementing the complete ZT aspect as recommended by NIST \cite{rose2020zero}, primarily focusing on using blockchain and miners for trust evaluation of incoming requests.

%Medical data is one of the most sensitive types of information that must be securely managed; typically, only accessible by authorized individuals, such as patients or their doctors. With the advent of IoT and 5G networks, medical records are increasingly available online, posing security risks like identity spoofing and unauthorized server access. 
Chen et al. proposed a ZT architecture for medical systems to address these challenges \cite{chen2020security}. Their approach goes beyond simple identity checks, introducing a four-dimensional (4D) access control framework assessing the `subject' (i.e., actor's identity), `object' (i.e., resource), `environment' (i.e., network or system), and `behavior' (i.e., access history). %By evaluating 'subject' and 'environment,' the system calculates risk scores for various combinations, informing trust assessments to determine resource access.  
Sultana et al. \cite{sultana2020towards} introduced a ZT blockchain approach for sharing medical test results. The system records the sender, receiver, and file location in blockchain blocks, focusing on transferring large data with a separate database. However, the proposed approach does not fully implement ZT aspects as recommended by NIST. %Senders undergo username/password authentication and device health checks, but receivers (e.g., patients) only require username/password and two-factor authentication. Additionally, 
The system lacks validation of peer behavior and dynamic privilege adjustments, which are key aspects of NIST's ZT architecture.  Dorri et al. \cite{dorri2017blockchain} used blockchain to secure IoT communication in smart homes. In their system, each smart home has a single central miner that controls access permissions based on identity, lacking the ZT property of evaluating device trustworthiness before granting access.  This check is, however, one of the main properties of a ZT architecture \cite{rose2020zero}. Nevertheless, there has not been any research that implements blockchain-based ZT in Edge-Cloud environments, which focuses on the applicability and feasibility aspects.

\section{Blockchain-based Zero Trust}
\label{section2}
\subsection{System Design and Processes}
\label{cap:system_design}

\begin{figure}[t]
    \centering
    \includegraphics[width=\linewidth]{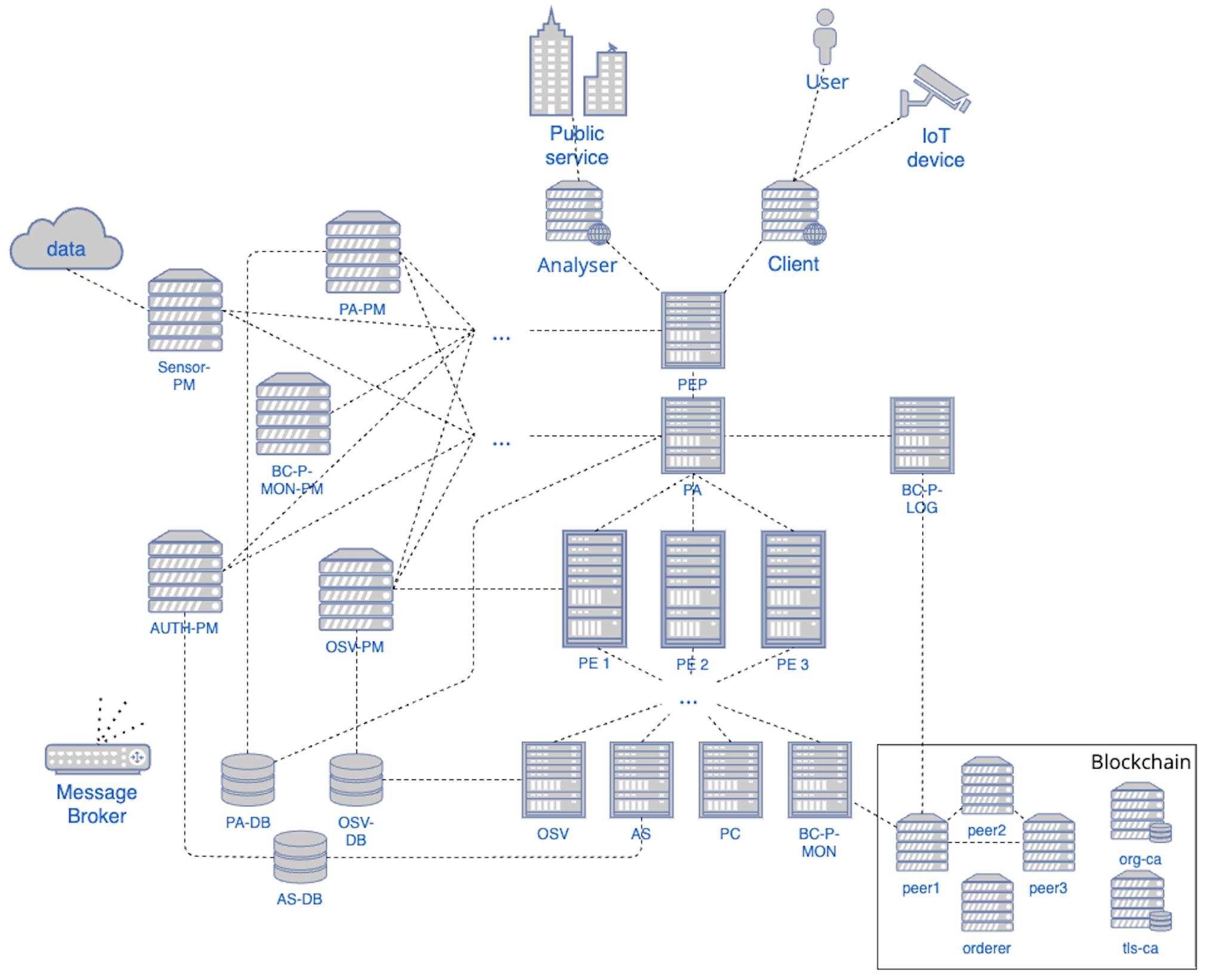}
    \caption{System components and network diagram.}
    \label{fig:big_picture}
\end{figure}

The developed system consists of three groups of components: (i) ZTA components, (ii) blockchain components, and (iii) IoT components. The system is designed to comply with the ZTA tenets and implement the core components mentioned in the ZTA definition of the NIST \cite{rose2020zero}. Furthermore, the blockchain components are implemented using the Hyperledger Fabric framework (HLF)\footnote{Hyperledger, https://www.hyperledger.org/}, building a permissioned blockchain. The PoC focuses on a smart city use case, addressing network properties like component distribution across multiple edge servers and low-computational-power IoT devices. The role of ZTA components is to enforce policies and grant or reject incoming requests. On the other side, the blockchain components include a distributed ledger implemented as a permissioned blockchain for logging incoming requests and their access decisions. The IoT components allow users and stationary IoT devices equipped with environmental sensors to access the system. The components interact by communicating over the Internet, with synchronous and asynchronous calls executed against REST APIs.  An overview of the system components and network diagram is depicted in Figure \ref{fig:big_picture}. In the following, we explain the role of each component and the processes within the developed system.

\subsection{ZTA Components} 

%\subsubsection{The Policy Enforcement Point (PEP)} PEP is the gateway from the outside world to the trust zone, and it communicates with the Policy Administrator (PA) and Policy Engines (PEs) to validate requests and make access decisions. The Trust Algorithm (TA) fetches data from various components to evaluate the authenticity and history of the requester and generate a decision. The Analyser component and the Client component are the two PEP client-side components used to interact with the system.% The Analyser component is an Angular application for monitoring the system's state, while the client component is a Python-based proxy server that intercepts requests and adds system-specific details.

\subsubsection{The Policy Enforcement Point (PEP)} PEP is a single logical component, but it can be divided into two physical components, as per the NIST definition \cite{rose2020zero}. In the PoC system, the PEP includes a server-side component, referred to as PEP, and two client-side components: Analyser and Client. Users and stationary IoT devices communicate through the Client component, while system administrators interact with the Analyser component. The Client component serves as a gateway, enhancing incoming requests with crucial actor details before forwarding them to the PEP. Access to the PEP component is limited to the two client-side components. Meanwhile, the Analyser component is accessible only to administrators and retrieves maintenance data, such as connected policy engines and actor request history. The PEP component receives incoming requests from the client-side components and forwards them to the Policy Administrator for validation. Upon successful validation and access granting, the PEP retrieves or sends the requested data to or from the relevant Persistence Managers. For example, if the local government deploys a new stationary device in a public park to measure outside temperature and wishes to connect it to the smart city network, an administrator would employ the Client component to submit a creation request to the PEP, which then forwards it to the PA. If the PA approves the access, the PEP dispatches the new stationary device's details to the Authentication Persistence Manager (AUTH-PM) for storage, making the device recognized by the system. In case of access denial, the PEP communicates no further with any PM but informs the requester of the rejection.

\subsubsection{The Policy Administrator (PA)} PA is responsible for validating incoming requests and generating access tokens for the PEP, which are necessary for PM access. The PA doesn't perform the actual validation but sends a validation request to all known PEs that conduct the verification. The validation process employs a straightforward consensus algorithm: all PEs begin validating the incoming request, and each PE informs the PA upon validation completion. If more than half of all PEs yield the same decision, it is accepted as correct. Depending on the request type, the PA either awaits PE validation completion or sends the validation request to all PEs and promptly notifies the PEP that validation has been initiated. Non-administrative data-saving requests are executed asynchronously, while administrative and GET requests are processed synchronously. This distinction improves system performance, especially when IoT devices regularly send sensor data, where response receipt is not always necessary. In the case of asynchronous execution, the PA informs the PEP through a message broker. Whether executed synchronously or asynchronously, the PA dispatches an access token to both the PEP and the corresponding PM upon access approval. This access token comprises a unique secret (string), a time-to-live value indicating its validity duration (in seconds), and a list of access rights specifying the type of requests permitted with the access token. The PEP must include this token with the request when accessing the PM.

\subsubsection{Policy Engine (PE)} The PoC employs multiple identical PEs, each running an instance of the Trust Algorithm (TA), which contains access policies and rules. The TA serves as the validation system's core and executes it as soon as a validation request arrives from the PA. In the PoC implementation, the TA is static in the order of security checks, which means that each incoming request from any actor is always validated the same way. All PEs are triggered simultaneously, and the TA executes synchronously. The TA collects various data about the requester (user or stationary actor) and the incoming request from different components. It then evaluates this data step by step to make a decision. The TA checks requester authenticity with data from the AS (identity checks), assesses requester vulnerabilities from the OSV component (environment checks), examines incoming request parameters using the PC (usage checks), and reviews requester history for suspicious activities from the Blockchain Peer Monitoring (BC-P-MON) component (behavior checks). The TA itself runs synchronously and is deterministic. All security checks are executed and the result of all of them are taken into consideration when building the validation decision. The caller is also informed about all security check failures and their severity. In our PoC implementation, for instance, we use the severity levels \texttt{LOW}, \texttt{MODERATE}, \texttt{HIGH}, and \texttt{CRITICAL}. All critical failures result in the rejection of the validated request. Once validation is complete, the PE transmits the decision to the PA. The TA validates incoming requests via the other components as explained in the next subsection.

\subsection{Other Components}
\subsubsection{Authentication Service (AS)}  The AS supplies data for the TA, containing information on known users and stationary actors, including their IDs, access rights, and IP/MAC addresses (only for stationary actors). The AS performs two types of checks: (a) for users, it verifies access rights, and (b) for stationary actors, it additionally compares incoming IP/MAC addresses with those in the database. The TA uses this data to ensure that the requester's access rights match the incoming request. The AS is a read-only component, unable to modify the database. To add, update, or delete authentication details, the Authentication Persistence Manager (AUTH-PM) is responsible (see the description of Persistence Managers (PMs) below for more PM responsibilities).

\subsubsection{Operating System Vulnerability (OSV)} The OSV is another component used for validation. It saves details about known vulnerabilities of operating systems. The source of this information could be an external service like the Common Vulnerabilities and Exposures (CVE) service\footnote{https://www.cve.org/} or the known vulnerabilities could be added manually via an API. For simplicity, some hardcoded vulnerabilities are inserted on startup, and new vulnerabilities can be added via the OSV API in the PoC. Details about the requester's operating system are included in the incoming request, and this information is checked against the data in the OSV's database.

\subsubsection{Parameter Checker (PC)} The PC is responsible for checking the parameters of incoming requests, more precisely, syntactic and semantic correctness of values, e.g., if an IP address is syntactically correct or if the temperature reading value is semantically valid.

\subsubsection{Blockchain Peer Monitoring (BC-P-MON)} The BC-P-MON component has the identity - i.e., certificate and private key - of a peer from within the blockchain network. It can fetch the historical data of actors from the blockchain through the installed chaincode (smart contract). The fetched data is checked for malicious or suspicious activities by the TA. For instance, if the last X requests had been rejected because the actor tried to fetch data from a restricted resource, and the next incoming request tries to fetch the same data again, the TA can recognize this suspicious activity and block the actor temporarily. There could be more sophisticated techniques when analyzing the history of actors in place. The TA could also check for specific patterns in the history to identify malicious or hijacked actors.

\subsubsection{Persistence Manager (PM)} There are many different PM components in the PoC. Each resource type has a dedicated PM in front of it; accessing it is only possible through the dedicated PM. Every incoming request must go through the whole ZT chain, starting with the client-side PEP component through the PA, PE, validation, and PM. No resource is allowed to take a shortcut. This implies that every request type has to have a PM to handle it. For instance, if the system wants to support reading the electricity consumption of public buildings, it has to implement a PM that can access the electricity data of those buildings. As seen in Figure \ref{fig:big_picture}, some PMs access databases are also accessed by validation components. For instance, the AUTH-PM and AS components are connected to the AS-DB. It is, however, not possible to modify data in the AS-DB from the AS component. The AS component only supports fetching data needed for validating incoming requests. It does not have methods to, e.g., delete or modify data, whereas the AUTH-PM component has full access (read and write) to the database. Additionally, as mentioned above, the PM can only be accessed via a valid access token, which must be registered by the PA first.

\subsubsection{Blockchain Peer Logging (BC-P-LOG)} This component is used for logging incoming actor requests in the blockchain. The PA sends a log request to this component after the PEs consent to a validation decision. The BC-P-LOG component also has the identity - again, certificate and private key - of a peer from the blockchain. It takes the incoming request and the decision outcome as input from the PA and sends it to the chaincode to persist in the blockchain.

\subsection{Blockchain} In addition to the above-mentioned components for ensuring ZT in the PoC, dedicated components are needed to build and run a permissioned blockchain within the system\footnote{The blockchain network is used as another additional component that provides input for the trust algorithm.}. The only connection between the ZT components and the blockchain components are the BC-P-* components. This connection is established by allowing those components to use the certificate and private key of the same peer from within the blockchain network to interact with the installed chaincode. All other ZT components do not know the blockchain components and vice versa.

For simplicity, the PoC only consists of a single organization (called \textit{org}), with its own certificate authority (called \textit{org-ca}). Apart from the organizations' certificate authority, there is also a certificate authority for creating TLS certificates (called \textit{tls-ca}). This TLS CA creates certificates for the organization itself, and the organization creates certificates for its peers. In the PoC, there are three peers (\textit{peer1}, \textit{peer2} and \textit{peer3}), but only one peer (\textit{peer1}) is used for submitting transaction proposals. This could be extended so that all peers are used for submitting transaction proposals so that the system can continue working when one peer is down. Furthermore, there is only one orderer node, the only component in the ordering service. Lastly, there is only one channel, and it only contains one chaincode. This chaincode contains smart contracts for fetching history data from actors - saved in the world state - and for persisting incoming requests of actors in the world state.

\begin{figure}
    \centering
    \includegraphics[width=\linewidth]{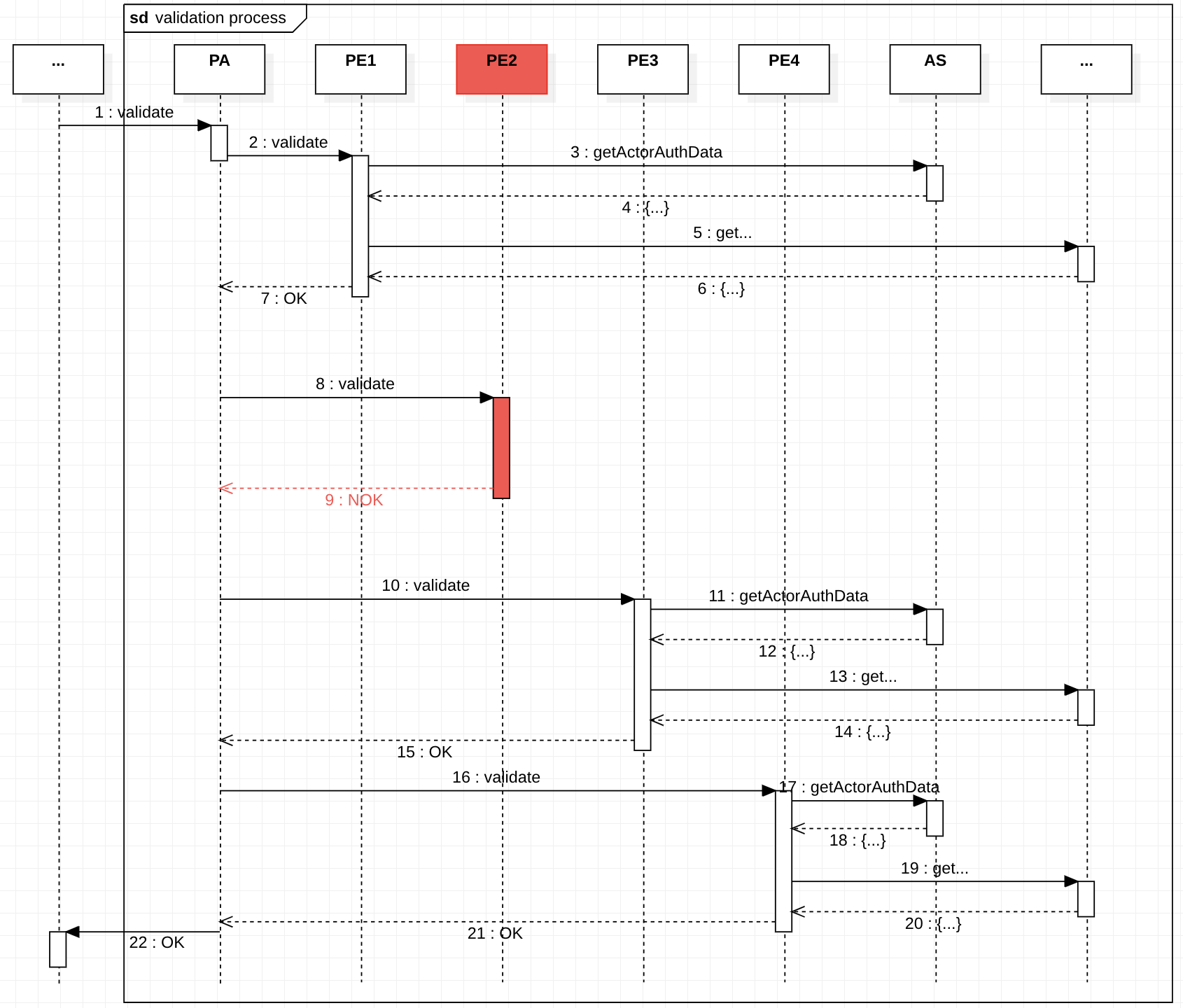}
    \caption{The validation process starts from the PA with compromised PE2 and acts maliciously.}
    \label{fig:hijacked_pe}
\end{figure}

In addition to the consensus algorithm of the blockchain, our PoC implementation uses another consensus algorithm in a different place as well. The validation process is also implemented to use a consensus algorithm. The consensus algorithm of choice is PBFT \cite{castro1999practical}. In the PoC, the PA lets all PEs validate the incoming request, and when the majority results in the same decision outcome, this result is taken as the correct decision. The Policy Administrator acts as the moderator, initiates the validation process, and waits for the results to accept the one with the most occurrences (\textit{Majority Voting} system). Figure \ref{fig:hijacked_pe} shows the validation process with a compromised PE. Again, for an attacker to control the validation process, it must control at least half of all PEs.

\subsection{Other components}
The remaining components (\textit{User}, \textit{IoT device}, and \textit{Public service}) are actual users and devices with sensors, respectively. A user could be an employee working for the local government (e.g., a gardener) who needs to read data from the smart city (e.g., humidity level in a park). This employee would use a smartphone or computer with the Python client component running. On top of that, a user-friendly UI application (e.g., a smartphone app or a website) that is connected to the Python client could be implemented. The public service office would use the Analyser UI directly for monitoring purposes. The identity of an actor is given in the HTTP request's "X-Requester" header. This information is in JSON format and has the following structure:

\begin{verbatim}
    X-Requester:
    {
        "agent": "...",
        "actor": "...",
        "ip_address": "...",
        "mac_address": "...",
        "os_id": "...",
        "os_version": "...",
        "auth_token": "..."
    }
\end{verbatim}

Actors must add their actor ID and their authentication token (API key) when accessing one of the client components. The Analyser component guarantees this by forcing users to provide their credentials (i.e., actor ID and API key) before opening the dashboard page. The Client component, on the other hand, has to be called with the X-Requester header already present in the HTTP header, but only with the credentials filled out, which will then be passed to the PEP component. However, both the Analyser and Client components will populate the remaining properties of the requester header - they don't have to and cannot be provided explicitly by the actors themselves. The X-Requester object is then moved from the header to the payload in the remaining API calls within \textcolor{black}{the trust zone \cite{rose2020zero}}, together with a description of the original incoming request of the actor. During the whole validation process, this payload is not modified.

\section{Evaluation}
\label{evaluation1}
In this section, we present implementation details and the evaluation results related to the non-functional properties of the PoC system. We do not discuss or rate actual computational logic like the TA or low-level system functionality but rather consider the system as a whole during the evaluation. We define specific test cases and execute them on the PoC implementation in a controlled environment to test performance and scalability. We implement various system variants for comparative purposes and subject them to the same test cases on the same testbed. In the following sections, we will discuss the evaluation process and results of the non-functional properties (i) {performance}, (ii)  {scalability}, and (iii) {implementation complexity}.

\subsection{Implementation details}
The ZTA-specific components are all Spring Boot Applications implemented in Java. Each component in the PoC (as presented in Figure \ref{fig:big_picture}) runs in a Docker container - except for the actual actors and users (e.g., public service) - and they are all orchestrated with Docker Compose\footnote{Docker, https://docs.docker.com/compose/}. All blockchain components use Docker images provided by HLF, all ZTA core components, validation components, PM components, and the BC logging component use Spring Boot Docker images. The HLF blockchain components had to be set up and initialized correctly. The ZTA components had been implemented from scratch in a microservice manner, meaning that the system is split up into its smallest possible units (microservices), and the communication between those units happens via REST APIs (synchronously). The client-side components of the PEP are implemented with Angular (\textit{Analyser} component) and Python (\textit{Client} component). For asynchronous communication, Redis\footnote{Redis, https://redis.io/} is used as the message broker of choice.

\subsection{Performance}
Evaluation of the PoC system's performance is most effectively done through comparison with alternative systems. We developed additional systems mirroring the PoC's functionality but with distinct architectures. These systems, alongside the PoC, were deployed in a test environment, and specific test cases were defined and executed. In the upcoming subsections, we detail the testbed, PoC variants, test cases, and their evaluation results.

\subsubsection{Testbed}
The PoC system consists of 30\texttt{+} components executed on a powerful server with Linux Ubuntu v22.04.2 (headless), 22 vCores, and 32 GB memory. %In a real-world scenario, the components would be split into multiple edge servers and devices, but to test the performance and scalability of the system, we decided to start everything up on a single, yet powerful server. 
%Our he hardware and software details about the server are presented in Table \ref{tab:server_specs}. 
The components were executed inside Docker containers which had the whole power of the host available.

%\begin{table}
 %   \caption{Software and hardware specifications of the test environment.}
  %  \centering
   % \begin{tabular}{|c|c|c|}
    %    \hline 
     %   Operating system & CPU & RAM \\ \hline
      %  Linux Ubuntu v22.04.2 (headless) & 22 vCores & 32 GB \\ \hline
    %\end{tabular}
    %\label{tab:server_specs}
%\end{table}

\subsubsection{Variants}
\begin{figure}[h]
  \centering
  \includegraphics[width=\linewidth]{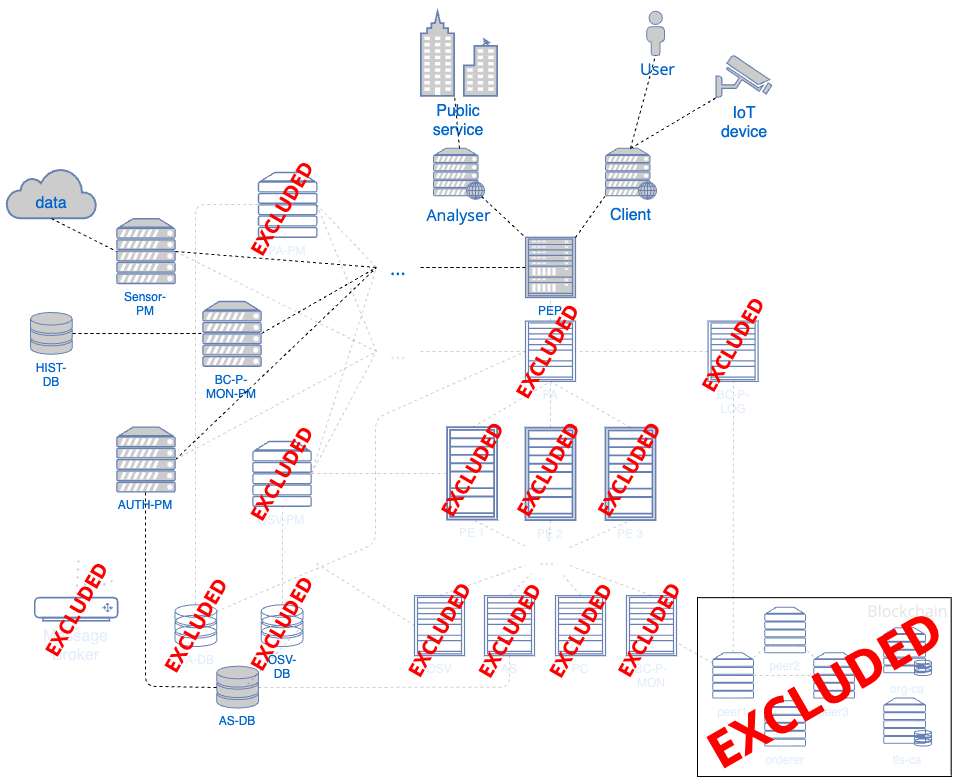}
  \caption{An overview of the "conventional" system. It excludes all blockchain components, the policy administrator, the policy engines and all validation components.}
  \label{fig:conv_variant}%
\end{figure}

\begin{figure}
    \centering
  \includegraphics[width=\linewidth]{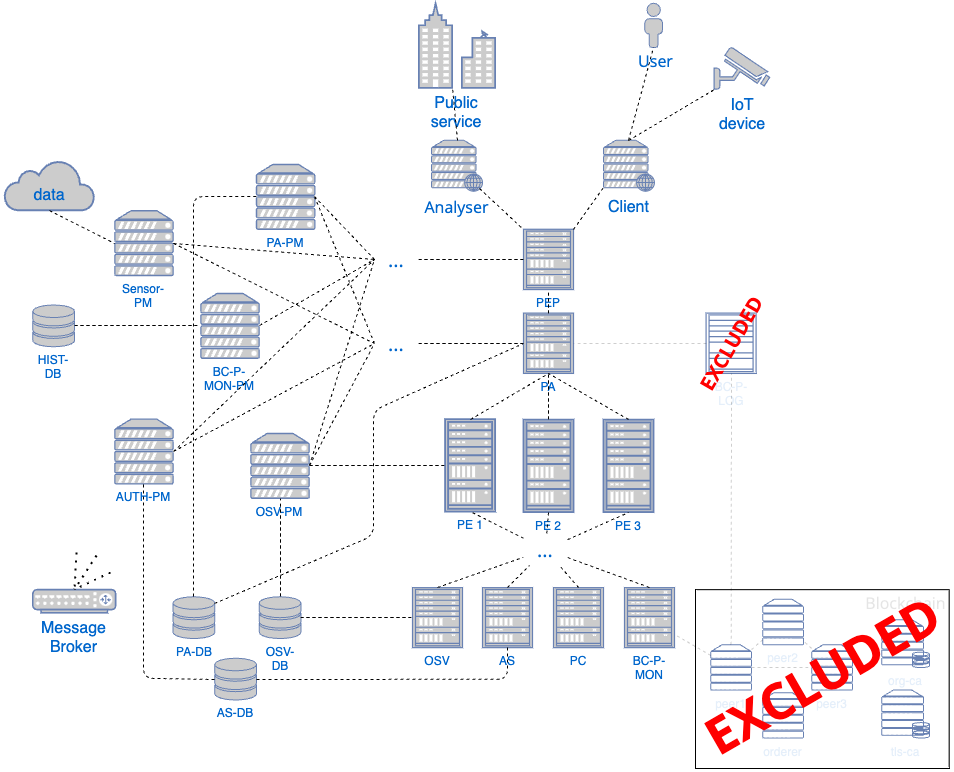}
  \caption{An overview of the "no-blockchain" system which excludes all blockchain elements and stores actor request history in a basic SQL database. This variant has yet another variant with 12 policy engines.}
  \label{fig:no_bc_variant}
\end{figure}

To have a comparable evaluation result, we implemented different variants of the PoC system. We implemented a basic system with no ZT properties or blockchain. The concrete variants are listed and described in Table \ref{tab:variants}, and depicted in Figure \ref{fig:conv_variant} and \ref{fig:no_bc_variant}. The variants are defined so that the main technologies of the PoC can be evaluated for performance impact, e.g., there are variants with ZT properties but no blockchain properties or variants with fewer and variants with more policy engine instances. This way, we can evaluate which part of the system slows the execution time more than others. All variants produce the same output when fed with the same inputs. This means that the functional properties are the same for all variants, and we can fully concentrate on evaluating the non-functional properties.

\begin{table}
    \caption{We tested five different variants relative to the original PoC system. In the table, the "excludes" column indicates components missing compared to the PoC system.}\label{tab:variants}
    \centering\resizebox{\linewidth}{!}{
    \begin{tabular}{|c|l|p{2.5cm}|p{2.4cm}|}
        \hline
        \textbf{Variant} & \textbf{Name} & \textbf{Description} & \textbf{Excludes} \\\hline
        1 & Conventional & Basic security mechanism where only authenticity and access rights are checked & All BC components, PE, PA, all validation components \\\hline
        2 & No BC & ZTA system but without a blockchain & All BC components \\\hline
        3 & No BC (x4) & Same as Variant 2, but with 4x more PEs (12 in total) & All BC components \\\hline
        4 & ZTA-BC & The original PoC & Nothing \\\hline
        5 & ZTA-BC (x4) & The original PoC, but with 4x more PEs (12 in total) & Nothing \\\hline
    \end{tabular}}
\end{table}

\subsubsection{Test cases}
The actual test cases were implemented as Python scripts, which were executed against the different variants. Every test script starts in the Client component, meaning each request goes through the whole process, from the Client component down to the database. We define five test cases, which all have different focus. Tests check the performance impact of synchronous requests compared to asynchronous requests, measuring how the number of policy engines affects the overall performance, giving insides into the scalability of a system variant, etc. The test cases and the actual requests executed in them are described in the following list. It needs to be mentioned that each test case sets itself up, meaning that users' required or stationary actors are initialized within the test case's process. The initialization steps are, however, excluded from the execution time.

\begin{itemize}
    \item \textbf{A user requests data that it does not have access to (TC1):} This is a straightforward test case to demonstrate which system is faster in recognizing that a user does not have permission to access the requested resource. It creates a user with insufficient access rights, and this user then executes a request to a resource for which it does not have access rights. The forbidden request is performed five times a row to eliminate any outliers because of server hiccups or other unexpected things. The execution time of this test case is the time between the sending of the first request and the response to the last request.
    \item \textbf{A stationary actor sends 20 temperature readings (POST). A user reads them afterward (TC2):} In this test case, a static actor is created, and this actor then sends 20 temperature readings to the system in a row - i.e., without halting between the requests. Immediately after, a user is created, which reads all tasks of this actor in a single read request. The execution time of this test case contains the 20 write requests only.
    \item \textbf{A stationary actor sends 1000 temperature readings (POST), and a user reads them afterward (TC3):} This test case is similar to the previous one (TC2), with the only difference being that the stationary actor sends 1000 temperature readings to the system. This test case can be compared with the following test case (TC4) to evaluate the performance difference between synchronous and asynchronous requests. The execution time in this test case only contains the execution time of the 1000 requests, i.e., the synchronous read request is not considered.
    \item \textbf{A user sends 1000 requests (GET) containing some temperature data (TC4):} This test case creates a stationary actor and a user. The static actor sends five temperature readings to persist, and the user reads the persisted data 1000 times (synchronous GET requests). Like the above test case, the execution time only contains 1000 requests.
    \item \textbf{Four stationary actors send data simultaneously (TC5):} This test case demonstrates how the system handles a high load from 4 simultaneously running threads. Each thread sends and reads data, and all threads execute the exact requests. Here, the execution time contains the whole execution from start to end, i.e., it includes the creation of the required actors and waits until all threads are finished.
\end{itemize}

\subsubsection{Performance evaluation}
\label{sec:performance_eval}
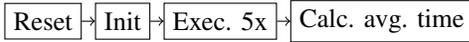
\begin{figure}
    \centering
    \begin{tikzpicture}[node distance=5pt]
        \node[rectangle, draw] (A) {Reset};
        \node[rectangle, draw, right = of A] (B) {Init};
        \node[rectangle, draw, right = of B] (C) {Exec. 5x};
        \node[rectangle, draw, right = of C] (D) {Calc. avg. time};
        \draw[->] (A) -- (B);
        \draw[->] (B) -- (C);
        \draw[->] (C) -- (D);
    \end{tikzpicture}
    \caption{Process for executing a test case against a system variant. This process is done for each pair of test cases and variants, i.e., 25 times in total.}
    \label{fig:eval_process}
\end{figure}

\begin{figure}
    % \centering
    \begin{tikzpicture}
    \begin{axis} [ybar, bar width=5pt, width=\linewidth, xlabel=Test case, xtick={0,1,2,3,4,5,6}, ymajorgrids, ymin=-150, tick label style={font=\tiny}, legend style={at={(0.05,0.8)},anchor=west}, nodes near coords, nodes near coords align={vertical}, nodes near coords style={font=\tiny}, nodes near coords style={anchor=east,rotate=90,font=\tiny,#1},legend cell align={left},legend style={font=\footnotesize}]
        \addplot coordinates {(1,0.741) (2,3.201) (3,143.140) (4,155.388) (5,331.083)};
        \addplot coordinates {(1,2.258) (2,4.180) (3,179.329) (4,490.825) (5,409.311)};
        \addplot coordinates {(1,5.636) (2,6.645) (3,262.942) (4,1174.277) (5,716.013)};
        \addplot coordinates {(1,2.407) (2,4.143) (3,180.462) (4,510.444) (5,416.399)};
        \addplot[brown,fill=burlywood] coordinates {(1,6.349) (2,6.635) (3,266.591) (4,1239.961) (5,721.760)};
        \legend{Conventional,No-BC,No-BC (x4),ZTA-BC,ZTA-BC (x4)}
    \end{axis}
    \end{tikzpicture}
    \caption{Execution times (in seconds) of the test cases against five different system setups: "Conventional" system (blue), ZTA without BC (red), ZTA without BC with 4x more PEs (lightbrown), ZTA with BC (grey), ZTA with BC with 4x more PEs (gold).}
    \label{fig:eval}
\end{figure}
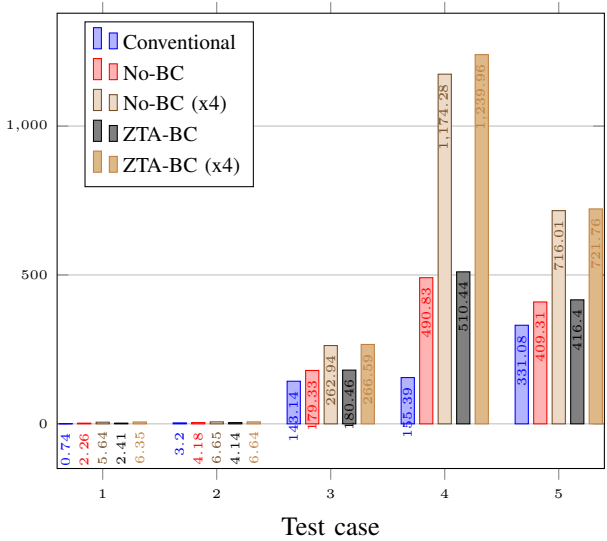

The above-listed test cases were executed against each above-mentioned system variant. The system had been pre-filled with some data before the tests were executed. Each test case had been executed five times in succession, and each system variant had been built and deployed from scratch before each test run - not before each test \textit{case}, but each test \textit{run}, i.e., before running the script for filling the system with initial data. This way, we can guarantee that every test is executed under the same conditions on each variant. The test cases are executed five times in a row without resetting the system in between because we also wanted to consider any increases in execution time with increasing data in the system. A running system is usually not empty, making the test cases more realistic. To evaluate the performance, the average execution time of the five executions is taken and compared between variants. The evaluation process is drawn in Figure \ref{fig:eval_process}. The performance assessment outcomes are illustrated in Figure \ref{fig:eval} from which several critical insights are derived and elaborated in the following paragraphs.
 
Read operations take much longer than write operations in our ZT system. Comparing test cases 3 and 4 between different variants, we see that in all ZT variants, TC3 (1000 \textit{read} requests) takes around \textbf{2x} the execution time compared to TC4 (1000 \textit{write} requests), only in the "conventional" system, the time is almost the same. The reason for the considerable time difference is apparent: read requests are validated synchronously, and write requests are asynchronous. Each read request waits for the PEs to reach consensus for all ZT variants, whereas write requests only trigger the background validation task immediately after returning to the requester. The conventional system does not encounter this difference because there is no validation process involved, i.e., read requests do not have to wait for consensus between policy engines because there are none.

More policy engines mean longer execution time - especially for read requests. As the validation process includes waiting for consensus between the policy engines, the execution time of a request validation increases the more policy engines are added to the system. The difference is, however, surprisingly not very big in all cases. For write requests, where the validation process is executed asynchronously, the execution time with 4 times more policy engines increases only around \textbf{2,4x} (TC1, ZTA-BC), \textbf{1,5x} (TC2, ZTA-BC), \textbf{1,5x} (TC3, ZTA-BC) and \textbf{1,7x} (TC5, ZTA-BC), respectively. This effect is more significant for read requests (synchronous validation): the increase in execution time in TC3 on the "ZTA-BC (x4)" in relation to TC4 on the "ZTA-BC (x4)" system is higher (approx. \textbf{\texttt{+}4.7x}) than in TC3 on the "ZTA-BC" in relation to TC4 on the "ZTA-BC" (approx. \textbf{\texttt{+}2.8x})

The presence of a blockchain does not significantly affect our ZT system's performance. When reading data from the blockchain, the query is executed \textit{without} consensus between the blockchain peers. This is the default behavior in the HLF. As the logging of actors' requests in the blockchain is executed asynchronously and fetching actors' request history from the blockchain does not go through a consensus process, the presence of a blockchain does not decrease the ZT system's performance significantly. For instance, the difference in execution time between the No-BC and ZTA-BC systems is around \textbf{\texttt{+}1.05x} (approx. \texttt{+}5\%) in all test cases.

The validation process takes the most time in our ZT system. This is verified empirically by checking the increase in execution time when extending the number of policy engines by multiple and, for instance, increasing the number of policy engines from 3 to 12 (4x) almost \textbf{triples} the execution time. This means that the increase in execution time is nearly linear to the increase in policy engines. Recall that the consensus algorithm in the request validation process asks all PEs for validation and only notifies the PEP about the validation result when more than half of the policy engines returned the same decision. This has the consequence that the more PEs are added to the system, the more decisions of PEs are needed to reach consensus, the longer it takes for the policy administrator to accept a decision, and the longer the overall execution time of the request takes.

\subsection{Scalability}
We have demonstrated the system's potential for scalability by theoretically allowing the addition of multiple PEs. However, in practice, when incorporating numerous components, such as PEPs, the PoC implementation experiences noticeable slowdowns. This doesn't necessarily indicate a lack of scalability in the system's design but can be attributed to various factors, including thread creation for asynchronous validation, hardware limitations, and context switches. Expanding system resources or adding validation properties to the trust algorithm is possible but requires implementation efforts and cannot be achieved dynamically during runtime with the current system design presented in this thesis. Adding new resources does not affect request execution times while augmenting trust algorithm policies impacts all incoming requests and slows down validation. %Scaling up actors involves installing the client-side policy enforcement point on their devices, either through downloads or pre-installed software. %Alternatively, it can be implemented on edge servers acting as gateways to the system. Scalability depends on specific use cases; for example, in the smart city scenario discussed here, all actors and users are known to the system, facilitating scalability. 
In conclusion, the PoC validates the feasibility and applicability of the system concept in smart city contexts but is not production-ready due to lacking certain non-functional properties (see Section \ref{sec:future_work}).

\subsection{Implementation complexity}
The PoC system comprises numerous components, including eight for zero-trust functionality (PA, PE 1-3, OSV, AS, PC, BC-P-MON) and six for hosting the permissioned blockchain. Each incoming request traverses various components, such as PEP, PA, multiple PEs, validation components, blockchain peers, orderer, PM, and finally, the database if the requester is trustworthy. The PoC involves 30 components, with each successfully validated request touching around 60\% of the system (19 components). This complexity arises due to the PoC's functional capabilities, involving many components that require intricate orchestration, especially when combining synchronous (REST APIs) and asynchronous (Message Broker - Pub/Sub) communication. Setting up the blockchain is also a complex task despite extensive documentation provided by the HLF. It involves multiple steps, including organizing, configuring certificate authorities, peers, and orderers, implementing encrypted communication (TLS), deploying chaincode, and developing client connections. Moreover, additional complexity arises from the validation consensus algorithm and the static nature of the current TA implementation.

\subsection{Limitations and future work}
\label{sec:future_work}
%This paper focused on designing and implementing a PoC system featuring ZT properties and an integrated blockchain component. 
The evaluation primarily addresses non-functional properties related to the system's design and implementation, complexity, scalability, and performance. The most apparent non-functional property of such systems, namely the \textit{security} property, is not evaluated as it is assumed that the security of a ZT system is, by design, more effective than the security in a perimeter-based system.  Note that the implemented PoC is not production-ready in its current state. The emphasis here was on demonstrating the system's feasibility, with specific non-functional properties, such as fault tolerance and high-availability. Performance can be boosted by implementing a more efficient consensus algorithm for validation, capable of handling multiple policy engines and high request loads. %For instance, the consensus algorithm could randomly select a subset of policy engines for verification instead of involving all of them. Another improvement avenue involves adding more policy administrators, managed by a load balancer to distribute workload evenly among policy engines. Currently, caching is underutilized in the system, although it has the potential to enhance performance significantly. However, careful consideration is needed to ensure caching doesn't compromise policy validation flexibility. Lastly, enhancing the policy enforcement process through machine learning is an option. Machine learning could be employed to recognize malicious behavior patterns and predict actor behavior, enhancing the system's ability to reject attackers or compromised actors effectively.

\section{Conclusion}\label{conclusion}
This paper introduced a novel framework mainly focused on designing and implementing an edge-supported system with a ZT architecture backed by blockchain. The proposed approach leverages blockchain as an immutable database to record and verify user requests, enhancing security by monitoring user activities for potential malicious behavior. Throughout this paper, we have elaborated on the framework's design, processes, and presented experimental results from a testbed, demonstrating its applicability in the context of smart cities. Our evaluation focuses on non-functional properties, including performance, scalability, and system complexity. However, it's important to note that this paper represents just an initial step towards the operationalization of the framework. In future work, we aim to provide a comprehensive technical framework encompassing technical and architectural aspects.

\section*{Acknowledgments}
Research has partially received funding from grant agreement Nos. 101079214 (AIoTwin) and 101070186 (TEADAL) by EU Horizon.


\bibliographystyle{IEEEtran}
\bibliography{main}

\begin{thebibliography}{10}
\providecommand{\url}[1]{#1}
\csname url@samestyle\endcsname
\providecommand{\newblock}{\relax}
\providecommand{\bibinfo}[2]{#2}
\providecommand{\BIBentrySTDinterwordspacing}{\spaceskip=0pt\relax}
\providecommand{\BIBentryALTinterwordstretchfactor}{4}
\providecommand{\BIBentryALTinterwordspacing}{\spaceskip=\fontdimen2\font plus
\BIBentryALTinterwordstretchfactor\fontdimen3\font minus \fontdimen4\font\relax}
\providecommand{\BIBforeignlanguage}[2]{{%
\expandafter\ifx\csname l@#1\endcsname\relax
\typeout{** WARNING: IEEEtran.bst: No hyphenation pattern has been}%
\typeout{** loaded for the language `#1'. Using the pattern for}%
\typeout{** the default language instead.}%
\else
\language=\csname l@#1\endcsname
\fi
#2}}
\providecommand{\BIBdecl}{\relax}
\BIBdecl

\bibitem{syed2022zero}
N.~F. Syed, S.~W. Shah, A.~Shaghaghi, A.~Anwar, Z.~Baig, and R.~Doss, ``Zero trust architecture (zta): A comprehensive survey,'' \emph{IEEE Access}, 2022.

\bibitem{sedlak2023privacy}
B.~Sedlak, I.~Murturi, P.~K. Donta, and S.~Dustdar, ``A privacy enforcing framework for data streams on the edge,'' \emph{IEEE Transactions on Emerging Topics in Computing}, 2023.

\bibitem{computers12100198}
P.~K. Donta, I.~Murturi, V.~Casamayor~Pujol, B.~Sedlak, and S.~Dustdar, ``Exploring the potential of distributed computing continuum systems,'' \emph{Computers}, vol.~12, no.~10, 2023.

\bibitem{rose2020zero}
S.~Rose, O.~Borchert, S.~Mitchell, and S.~Connelly, ``Zero trust architecture,'' National Institute of Standards and Technology, Tech. Rep., 2020.

\bibitem{gharaibeh2017smart}
A.~Gharaibeh, M.~A. Salahuddin, S.~J. Hussini, A.~Khreishah, I.~Khalil, M.~Guizani, and A.~Al-Fuqaha, ``Smart cities: A survey on data management, security, and enabling technologies,'' \emph{IEEE Communications Surveys \& Tutorials}, vol.~19, no.~4, pp. 2456--2501, 2017.

\bibitem{salman2018security}
T.~Salman, M.~Zolanvari, A.~Erbad, R.~Jain, and M.~Samaka, ``Security services using blockchains: A state of the art survey,'' \emph{IEEE communications surveys \& tutorials}, vol.~21, no.~1, pp. 858--880, 2018.

\bibitem{xiaojian2021power}
Z.~Xiaojian, C.~Liandong, F.~Jie, W.~Xiangqun, and W.~Qi, ``Power iot security protection architecture based on zero trust framework,'' in \emph{2021 IEEE 5th International Conference on Cryptography, Security and Privacy (CSP)}.\hskip 1em plus 0.5em minus 0.4em\relax IEEE, 2021, pp. 166--170.

\bibitem{decusatis2016implementing}
C.~DeCusatis, P.~Liengtiraphan, A.~Sager, and M.~Pinelli, ``Implementing zero trust cloud networks with transport access control and first packet authentication,'' in \emph{2016 IEEE International Conference on Smart Cloud (SmartCloud)}.\hskip 1em plus 0.5em minus 0.4em\relax IEEE, 2016, pp. 5--10.

\bibitem{samaniego2018zero}
M.~Samaniego and R.~Deters, ``Zero-trust hierarchical management in iot,'' in \emph{2018 IEEE international congress on Internet of Things (ICIOT)}.\hskip 1em plus 0.5em minus 0.4em\relax IEEE, 2018, pp. 88--95.

\bibitem{chen2020security}
B.~Chen, S.~Qiao, J.~Zhao, D.~Liu, X.~Shi, M.~Lyu, H.~Chen, H.~Lu, and Y.~Zhai, ``A security awareness and protection system for 5g smart healthcare based on zero-trust architecture,'' \emph{IEEE Internet of Things Journal}, vol.~8, no.~13, pp. 10\,248--10\,263, 2020.

\bibitem{sultana2020towards}
M.~Sultana, A.~Hossain, F.~Laila, K.~A. Taher, and M.~N. Islam, ``Towards developing a secure medical image sharing system based on zero trust principles and blockchain technology,'' \emph{BMC Medical Informatics and Decision Making}, vol.~20, no.~1, pp. 1--10, 2020.

\bibitem{dorri2017blockchain}
A.~Dorri, S.~S. Kanhere, R.~Jurdak, and P.~Gauravaram, ``Blockchain for iot security and privacy: The case study of a smart home,'' in \emph{2017 IEEE international conference on pervasive computing and communications workshops (PerCom workshops)}.\hskip 1em plus 0.5em minus 0.4em\relax IEEE, 2017, pp. 618--623.

\bibitem{castro1999practical}
M.~Castro, B.~Liskov \emph{et~al.}, ``Practical byzantine fault tolerance,'' in \emph{OsDI}, vol.~99, no. 1999, 1999, pp. 173--186.

\end{thebibliography}

\vfill

\end{document}